# Probing Purcell enhancement and photon collection efficiency of InAs quantum dots at nodes of the cavity electric field

Matthew Jordan [1,2], Petros Androvitsaneas [1,2], Rachel N Clark [1,2], Aristotelis Trapalis [3], Ian Farrer [3,4], Wolfgang Langbein [5], and Anthony J. Bennett [1,2,5,*]

[1] School of Engineering, Cardiff University, Queen's Buildings, The Parade, Cardiff, CF24 3AA, United Kingdom

[2] Translational Research Hub, Maindy Road, Cardiff, CF24 4HQ, United Kingdom

[3] Department of Electronic and Electrical Engineering, University of Sheffield, Mappin Steet, Sheffield, S1 3JD, United Kingdom

[4] EPSRC National Epitaxy Facility, University of Sheffield, Broad Lane, Sheffield, S3 7HQ, United Kingdom

[5] School of Physics and Astronomy, Cardiff University, Queen's Buildings, The Parade, Cardiff, CF24 3AA, United Kingdom

* Correspondence to: BennettA19@cardiff.ac.uk

The interaction of excitonic transitions with confined photonic modes enables tests of quantum physics and design of efficient optoelectronic devices. Here we study how key metrics such as Purcell factor, $\beta$-factor and collection efficiency are determined by the non-cavity modes which exist in real devices, taking the well-studied micropillar cavity as an example. Samples with dots at different positions in the cavity field allow us to quantify the effect of the non-cavity modes and show that the zero-phonon line and the phonon-assisted emission into the cavity mode $HE_{11}$ is suppressed by positioning dots at the field node.

Single-photon sources are crucial to applications in quantum metrology [1], secure quantum communications [2], and optical quantum computing [3,4]. In solid state devices the local photonic environment can be structured to promote the efficient collection of photons into a lens. This can be achieved by suppressing emission into unwanted directions, such as in a photonic crystal [5,6], or by promoting emission into a single mode that couples well to the far field optics, such as with a nano-antenna [7,8]. Numerical design of these structures often focuses on the localized 'cavity' modes of high quality factor because these modes show a clear initial decay and can be calculated using a small simulation volume, and thus in a practical run-time. Simulations are less able to predict non-cavity (often called 'leaky') modes that are spectrally broad, overlapping, and are difficult to extract from numerical FDTD and FEM simulations. Understanding the role of these non-cavity decay channels is essential for a complete understanding of photon source behavior, as they provide alternative radiative decay channels.

A popular design for efficiently generating single photons embeds semiconductor quantum dots (QDs) in a monolithic micropillar cavity [9-11]. A cavity spacer layer between distributed Bragg reflectors (DBRs) forms cavity modes within the stopband of the DBRs, and the lateral mode is confined by etching the planar structure into pillars. QDs are typically positioned at an anti-node in the cavity field to maximize coupling to the fundamental $HE_{11}$ cavity mode [12,13]. Recent years have seen steady progress towards unity efficiency, single-photon purity and indistinguishability in such sources [14-18] driven by advances in processing [14], in-situ lithography [19], and coherent excitation [20,21].

In the "weak coupling" regime, where coupling strength is well below the mode or emitter linewidth, a transition's emission rate is increased on resonance via Purcell enhancement [9,22], leading to reduced emission into non-cavity modes [23]. The coupling between a single transition and the cavity mode at zero detuning is quantified by the Purcell factor $F_p$, typically defined as the ratio of the radiative decay rate inside the cavity, given by the sum of the decay into the localized cavity mode $\Gamma_c$ and the decay into the non-cavity, leaky modes $\Gamma_L$ [24], relative to the radiative decay rate in a uniform homogenous medium $\Gamma_0$. In the case where $\Gamma_c \gg \Gamma_L$ the Purcell factor can also be approximated analytically using the mode volume $V$, quality factor $Q$, and effective refractive index $n_{\text{eff}}$ [25]:

$$F_p = \frac{\Gamma_c + \Gamma_L}{\Gamma_0} \approx \frac{3Q\lambda^3}{4\pi^2 V n_{\text{eff}}^3} \quad (1)$$

The fraction of photons emitted into the cavity mode is called the spontaneous emission (SE) coupling factor, or $\beta$ factor. It is thus related to the Purcell factor by:

$$\beta = \frac{\Gamma_C}{\Gamma_C + \Gamma_L} = \frac{F_P}{F_P + \Gamma_L/\Gamma_o} \quad (2)$$

Multiple methods exist for experimentally determining $F_p$ [26], but the most direct is to measure the decay of the emission intensity in time. In solid state systems, it is not always possible to make a comparable measurement of the decay $\Gamma_0$ for the same source, and therefore authors estimate $\Gamma_0$ from an emitter assumed to be comparable in another sample [27]. Alternatively, the transition-mode detuning can be varied by temperature [26,28], deposition [29], static electric [30] or magnetic [31] fields. Besides the fact that these control parameters may lead to a change in $\Gamma_0$ and Q to some extent, there is an implicit assumption that a decay rate detuned from a mode, $\Gamma_L$, is comparable to that in a homogenous medium, $\Gamma_0$. However, this is generally not the case, as we show here, because the inhomogeneous dielectric environment and its influence on the local photon density of states (LDOS) remains. We note that the notion of emission into individual modes is approximate, as this LDOS should be expressed as a sum over all modes of the system and interference occurs between modes [32].

Here we investigate the effect of the non-cavity modes, comparing a pair of nominally identical micropillar samples, differing only in that one has QDs located at a field anti-node where Purcell-enhanced emission into $HE_{11}$ dominates, and the other has QDs at a node, where coupling into $HE_{11}$ is suppressed. We compare the emission of QD transitions in both cases averaged over several cavities to quantify the emission that is collected from non-cavity modes. Furthermore, individual cavities with a small number of well-resolved transitions are studied to investigate phonon-assisted emission into $HE_{11}$. Single transitions are tuned across the cavity mode using a Faraday-geometry magnetic field, to determine the change in decay time arising from the Purcell effect. This study provides a toolset to quantify the effects of non-radiative decay, non-cavity modes, and phonon-assisted coupling in this well-studied system.

We begin by simulating 1.85 µm diameter circular GaAs-$Al_{0.95}Ga_{0.05}As$ micropillars in ANSYS Lumerical FDTD, as shown in Fig. 1a. The micropillar consists of a 267.9 nm GaAs spacer layer, with DBRs of 17 λ/4 alternating GaAs and $Al_{0.95}Ga_{0.05}As$ pairs above and 26 pairs below, standing on a planar GaAs substrate. The refractive indices of the materials include dispersion but are assumed to have no imaginary component. QD emission is modelled as an electric dipole source at $x=y=0$, oriented along the $x$-axis. The height of the dipole ($z$) is varied to probe nodes and antinodes of the mode as sketched in Fig. 1b. The dipole is driven with a Gaussian pulse with 5.6 ps full width at half maximum (FWHM), covering a spectral range from 840 nm to 1070 nm. A total spectral average power of 3.79 fW is injected by the dipole. The simulation volume has $x$ and $y$ dimensions of 5.5 µm and a $z$ dimension of 10 µm, with perfectly matched layers at the simulation boundaries. Each surface of the simulation volume has a planar frequency-domain field monitor to record the electric field and transmission. The spectrally-resolved power through the top surface monitor reveals a $HE_{11}$ quality factor of ~21600, consistent with analytical models [33]. Additional planar field monitors normal to each dimension $x$, $y$, and $z$ intersect the origin. Simulations ran until the system energy dropped to $10^{-6}$ of its initial value.

The electric field amplitude at the $HE_{11}$ mode wavelength of 937.827 nm on the $y=0$ ($x$-$z$) and $x=0$ ($y$-$z$) planes are shown in Fig. 1c-f. For a dipole at the anti-node of the electric field ($z=0$), the distribution on the plane $y=0$ parallel to the dipole reveals a dominant component of emission in the vertical direction (Fig. 1c). The distribution on the plane $x=0$, orthogonal to the dipole (Fig. 1d) also shows strong emission in the vertical direction, but also some guiding in the horizontal direction at the spacer layer, which will not be collected by an optic above the sample. This is not seen on the plane parallel to the dipole, which cannot emit along its axis. Conversely, when the dipole is placed at the lower node of the cavity electric field, at one quarter of the spacer height ($z=-0.067$ µm), the field intensity does not become enhanced and guided modes of high radial quantum number dominate. A large fraction of the light escapes downward into the high index substrate.

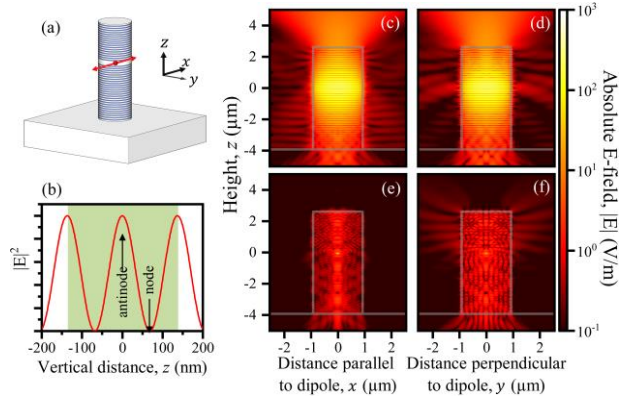

FIG. 1. (a) Simulation geometry containing a micropillar with 17 (26) upper (lower) DBR pairs, excited by a dipole orientated along the $x$-axis. (b) Sketch of the standing wave in the cavity mode electric field intensity in the GaAs spacer (green). (c-f) Absolute electric field |E| cross-sectional profiles for a 1.85 µm micropillar (grey outline) at the $HE_{11}$ mode wavelength. (c,d) for a dipole at the anti-node position in the center of the spacer layer: (c) on the $y=0$ plane ($x$-$z$), parallel to the dipole and (d) on the $x=0$ plane ($y$-$z$), perpendicular to the dipole. (e,f) as (c,d), but for a dipole at the lower node.

To quantify the direction of emission, we define an outcoupling efficiency parameter ξ, describing the fraction of the total power integrated over all directions $\Gamma_C + \Gamma_L$ that is directed towards the top of the micropillar via the cavity mode $\Gamma_1$:

$$\xi = \frac{\Gamma_1}{\Gamma_C + \Gamma_L} \quad (3)$$

To optimize collection efficiency it is thus desired to maximize this value [34].

In Fig. 2a, we show the spontaneous-emission coupling factor $\beta$, outcoupling efficiency ξ, and Purcell factor $F_P$ as functions of the dipole position $z$ within the spacer. $\beta$ was calculated using Equation 2 based on the energy-resolved total transmission through the closed volume around the simulation volume, isolating the $HE_{11}$ mode from the spectrally broad non-cavity-mode emission by fitting a Lorentzian function. We find $\beta = 0.989$ when the dipole is at the anti-nodes in the bottom, middle, and top of the spacer, and $\beta = 5 \times 10^{-4}$ at the nodes. The outcoupling efficiency ξ is determined using the ratio of the energy-resolved emission through the top planar monitor, again isolating the $HE_{11}$ mode from the background using a Lorentzian function, to the total emissions in all directions, as per Equation 3. ξ reaches 0.849 at the anti-node and falls to $3 \times 10^{-6}$ at the nodes.

$F_P$ was also calculated from the total power emitted in all directions at the $HE_{11}$ mode energy, normalized to the dipole source power in a homogenous medium of the local refractive index. It closely follows a sinusoidal pattern peaking at $F_P \sim 78$ at anti-node positions. Conversely, $F_P$ falls to 0.91 at the lower node, indicating some suppression of radiative emission at the mode wavelength at these source heights. It should be noted that the FDTD simulations do not simulate rough sidewalls on the cavities, which have been reported [10-12] to dominate losses for cavity diameters below 2.5 µm, suppressing $Q$ and pushing $F_P$ closer to unity.

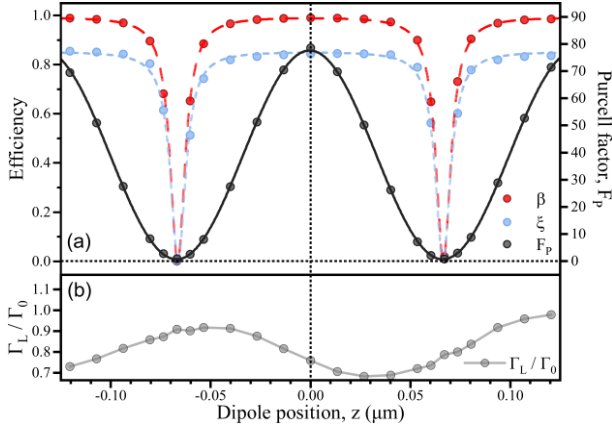

FIG. 2. Simulation results showing how (a) spontaneous-emission enhancement factor $\beta$ (red), outcoupling efficiency $\xi$ (blue), and Purcell factor $F_P$ (black), and (b) the ratio of non-cavity modes decay rate to the decay rate in a homogenous medium $\Gamma_L/\Gamma_0$ vary with dipole source height z.

Fig. 2b shows the emission via non-cavity modes relative to the emission in a homogenous medium, $\Gamma_L/\Gamma_0$, calculated from $F_P$ and $\beta$ as a function of $z$. This also shows a near-sinusoidal pattern but with a different phase and longer period than the variation in $F_p$ shown in Fig. 2a. We hypothesize that this weak variation in non-cavity modes is a result of reflections from the semiconductor-air interfaces at the surface of the pillar. This difference in period also clearly shows that the cavity mode has no direct influence on the non-cavity modes and there is no conservation in LDOS at a given frequency. We note that at all heights the $\Gamma_L/\Gamma_0$ is below unity, being 0.76 at position $z = 0$ where radiative emission is suppressed. Additionally, because the spatial distribution of the non-cavity modes is independent of the $HE_{11}$ they will have a different effective index, and consequently different periodicity in Fig. 2b.

In the following experimental work, we study samples with emitters at different heights, nominally identical to the simulated structures. Using direct-write projection photolithography, two samples grown on 3-in. wafers by molecular beam epitaxy were etched into cylindrical micropillars of a range of sizes [11], although here we focus on those of 1.85 µm diameter as simulated. Sample A contains InAs QDs at the mid-point of the cavity spacer, at a cavity mode anti-node. Conversely, sample B contains InAs QDs at one-quarter of the spacer height, at a node of the cavity mode. The cavity mode energy, measured for a planar structure, varies between the samples by 0.73%. Therefore, the following data is plotted against energy relative to the cavity mode determined from white light reflectivity spectra at 4 K.

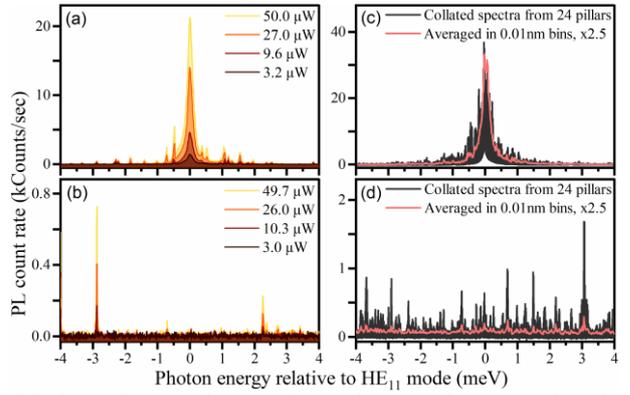

FIG. 3. PL spectra integrated over 10 s at varying powers for 1.85 µm pillars without transitions resonant with the $HE_{11}$ mode in (a) Sample A and (b) Sample B. Averaged spectra from twenty-four 1.85 µm pillars at low power in (c) sample A and (d) sample B.

Photoluminescence (PL) spectra under 850 nm CW laser excitation were recorded from single micropillars using a confocal arrangement with an objective lens of numerical aperture NA = 0.81. Exemplary cavities, shown in Fig. 3a and b were selected to show no sharp transitions near the $HE_{11}$ cavity mode at 4 K and 50 µW excitation power. Sample A displays an emission peak at the $HE_{11}$ mode energy, as if driven by a spectrally broad internal light source in the cavity, such that the intensity at the mode energy is a factor of 418 ± 48 greater than when far from the mode. This "cavity feeding" is the result of cavity-enhanced phonon-assisted emission by spectrally-detuned transitions within the cavity [35,36]. The intensity at the mode energy increases and saturates at higher powers, as the zero-phonon transitions also do. Conversely, when sample B is exposed to identical excitation conditions there is no visible mode, confirming that positioning the dots at a node suppresses the cavity-enhanced phonon-assisted emission. This is consistent with the simulations in Fig. 2 for a dipole at the node. The ratio of intensities at the mode energy for the two cavities is 764 ± 70, suggesting that even if emission is reaching the collection optic via the non-cavity modes, it does not display any spectral structure within the 35 nm range acquired with the spectrometer, and is strongly suppressed relative to photon collection via the $HE_{11}$ mode in Sample A.

It is non-trivial to compare two individual cavities in any experiment where the x-y position of the dot, its $\Gamma_0$, and photo-physics, are unknown. We therefore also probe, under identical conditions, twenty-four neighboring micropillars of ~1.85 µm diameter at 4 K under 50 µW power. For each cavity, the spectrum is offset by the mode energy determined from a reflectivity measurement to produce, in Fig. 3c,d, plots of the mean spectrum. Sample A shows an intensity enhancement for transitions near the mode. We estimate a Q factor of ~5000 in PL, and an enhancement of the photon collection by a factor of 560 ± 150 over the transitions spectrally detuned from the mode. Conversely, in Sample B, there is no enhancement at the mode energy; the transitions' brightness displays no trend with energy. Intensity at the mode energy is a factor of 395 ± 5 below the comparable value in sample A. Given that the intensity of light collected via the cavity mode is approximated by $\xi$, one might expect the ratio of intensities between the two samples to be below $10^{-5}$. One explanation is that in sample B the majority of the light collected is scattered from the rough pillar sidewalls and this is collected ~395 times less efficiently than the $HE_{11}$ mode in sample A.

In order to tune a given QD transition relative to the cavity mode, we performed measurements under a magnetic field $B$ applied along the $z$-axis. In this Faraday geometry the Zeeman effect and diamagnetic response tunes the transition energy relative to the cavity mode, as shown in the insert to Fig. 4a. It is assumed that the magnetic field does not modify the $\Gamma_L$ and $\Gamma_0$ emission rates. A 76 MHz pulsed laser and avalanche photodiodes (APDs) of 247 ps Gaussian FWHM timing jitter were used to record the decay time of each transition at a time-averaged power of 14.4 µW, where the transition is substantially below saturation. The data was collected from pre-selected bright, isolated transitions less than 1 nm in the wavelength above the cavity resonance at $B = 0$. Most transitions studied are in separate micropillars, except B3a and B3b. The Zeeman energy splitting increases linearly, consistent with excitonic $g$-factors between 2.639 and 3.525, as would be expected for InAs QDs of this type [37,38]. There is also a quadratic diamagnetic blueshift with coefficients ranging from 6.67 to 9.66 µeV/T$^2$.

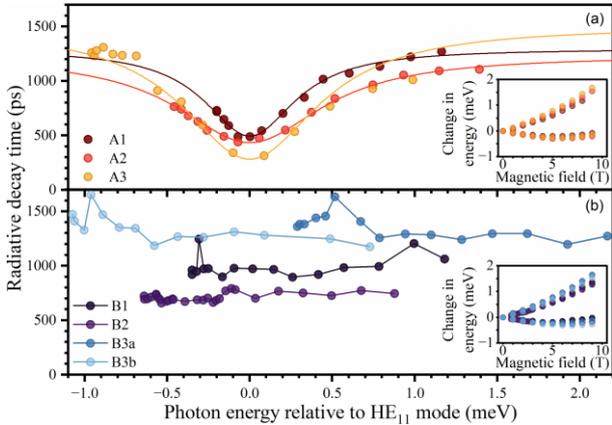

FIG. 4. Radiative decay time as a function of transition-mode detuning via magnetic field tuning in Faraday geometry for (a) sample A and (b) B. Different transitions are indicated in different colors, as labeled. Insets show the dependence of the transitions' energy as function of magnetic field.

In sample A (Fig. 4a) we observe a reduction of the decay time for transitions resonant with the cavity mode, consistent with Purcell enhancement. The data has been fitted with a Lorentzian curve, which matches the data in most cases, but some asymmetric deviation far from the mode is observed for A3. The ratio of the lifetime at large detunings to the minimum lifetime, for transitions A1, A2 and A3, is $2.53 \pm 0.05$, $3.80 \pm 0.09$, and $5.02 \pm 0.69$ respectively. Noting that as the lifetime at large detuning is determined by $\Gamma_L$, $F_P$ as defined in Eq. 1 will be a factor of $\Gamma_L/\Gamma_0 = 0.76$ lower than these ratios. Corresponding $\beta$ values of $0.72 \pm 0.02$, $0.79 \pm 0.04$, and $0.83 \pm 0.16$ are found using Eq. 2.

In contrast, for sample B there does not appear to be any cavity-mode-dependent enhancement or suppression of the lifetime of a quantum dot transition, confirming that the coupling of the transition to the cavity mode is strongly suppressed. The minimal variation in lifetime also confirms the LDOS given by the sum of the non-cavity modes is spectrally flat over the tuning range of a few micro-eV. In practical terms it will be hard to engineer a sample with the several 100 nanometer wavelength variation required to probe wide-band spectral structure of the non-cavity modes, which is broader than the stop-band of the mirrors. The reduced density of states at this location in the cavity, $\Gamma_L/\Gamma_0 = 0.91$ found in the simulations (Fig. 2b), implies a small Purcell suppression of radiative decay for all these transitions.

$F_P$ and $\beta$ are lower than predicted, which we attribute to the differences between the ideal case modelled and a real quantum dot in a micropillar. Firstly, the quality factors of the micropillars studied experimentally are significantly lower than those simulated, which we attribute to the roughness of the sidewalls. This roughness can be quantified by a single value that benchmarks how cavity loss is affected by scattering, known as the sidewall loss coefficient, previously shown to be $k_s = 50 \pm 20$ pm in these samples [11]. Assuming roughness leaves the mode volume of the micropillar unchanged, one might expect a proportional decrease in $F_P$, based on Equation 1. The micropillars in each sample displayed Q ~ 5000, a factor of 4.32 below the simulated value, which would reduce the maximum achievable $F_P$ to 18. Secondly, the QDs may not be located on the central axis of the pillar. Assuming a uniform density of QDs over the cross-sectional area of the pillar, we can estimate a median radial displacement of 0.654 µm. Simulations of 1.85 µm pillars with dipole sources displaced from $x=y=0$ indicate that the Purcell factor decreases rapidly with displacement from the center of the micropillar in a Gaussian-like pattern, such that we might expect the median $F_P$ to be further reduced to 5.83. A slightly larger radial displacement of > 0.7 µm would be sufficient to reduce $F_p$ to the experimentally observed values.

In the cavity we have considered here, the LDOS of the non-cavity modes results in $\Gamma_L/\Gamma_0 = 0.68$-$0.98$ as the dipole position in varied along $z$. This implies the common experimental practice of tuning transitions relative to the cavity mode systematically overestimates the Purcell factor, because the system moves from enhanced decay rate on resonance to suppression of emission at large detuning. Using a sample with emitters at a node in the cavity mode, we have confirmed experimentally that the non-cavity modes are spectrally broad relative to the limited tuning range we can access. Further simulations would be required to determine the broadband structure of the non-cavity modes LDOS. Also, future work may focus on the design of cavities with modified mirrors, spacer and diameter that result in a reduced $\Gamma_L/\Gamma_0$ over an increased volume in the cavity, to increase $F_p$ and $\beta$ for a larger number of dots. This study underlines the importance of a proper understanding of the interplay between cavity and non-cavity modes in experimental determination of the Purcell factor, which has implications for the characterization of photon sources in all nanostructured systems including nano-lasers, photonic crystals, open cavities, nano-antennae and integrated photonics.

We acknowledge financial support from EPSRC Grant No. EP/T017813/1 and EP/T001062/1. Sample processing was carried out in the cleanroom of the ERDF funded Institute for Compound Semiconductors at Cardiff University. RC was supported by grant EP/S024441/1 and the National Physical Laboratory.